\documentclass[iop,preprint,10pt]{emulateapj}
\usepackage[]{natbib}
\usepackage{graphicx}
\usepackage{subfigure}
\usepackage[dvipsnames]{color}
\usepackage{float}
\usepackage{amsmath}
\usepackage{fancyref}
\definecolor{orange}{RGB}{255,69,0}
\definecolor{green}{RGB}{0,255,0}
\definecolor{darkred}{RGB}{139,0,0}
\usepackage{amssymb}
\usepackage{lineno}
\usepackage[colorlinks=true,
            linkcolor=orange,
            urlcolor=magenta,
            citecolor=blue]{hyperref}
\usepackage{nameref}
\usepackage{array,multirow,makecell}
\usepackage{tabularx}

\usepackage{morefloats}
\usepackage{gensymb}

\begin{document}
\title
{Multi-frequency variability study of Ton 599 during high activity of 2017  }

\author{Raj Prince$^{1}$}
\affil{$^{1}$Raman Research Institute, Sadashivanagar, Bangalore 560080, India \\
        } 

\email{rajprince@rri.res.in}
\begin{abstract}
In this work, I have presented a multi-frequency variability and correlation study of the blazar Ton 599, which was observed first time in flaring
state at the end of 2017. Data from \textit{Fermi}-LAT, Swift-XRT/UVOT, Steward Observatory, and OVRO (15 GHz) is used, and it is found that
the source is more variable in $\gamma$-ray and optical/UV than X-ray and radio.
Large variations in the degree of polarization (DoP) and position angle (PA) is noticed during the flaring period. Maximum flux during $\gamma$-ray flare is found to be 12.63$\times$10$^{-7}$ at MJD 58057.5 from the 1-day bin light curve (LC), 
which is the maximum flux ever achieved by this source. It is further found that all the peaks of flare are very symmetric, which suggests
the cooling time of electrons is much smaller than light crossing time. Using 1-day as a fast variability time, 
the size of the $\gamma$-ray emission region is estimated as 1.88$\times$10$^{16}$ cm. Two 42 GeV of photons are detected during the flare which puts a
constraint on the location of the emission region, and it is found that the $\gamma$-ray emitting blob is located at the outer edge or outside the broad line region (BLR). 
A trend of increasing fractional variability towards higher energies is also seen. Strong correlations were seen between $\gamma$-ray, optical/UV,
X-ray, and radio (15 GHz) emission. A small time lag between $\gamma$-ray and optical/UV suggest their emission to be co-spatial while lag of 27 
days between $\gamma$-ray and OVRO (15 GHz) suggest two different emission zone separated by a distance of $\sim$ 5 pc.

\end{abstract}
\keywords{galaxies: active; gamma rays: galaxies; individuals: Ton 599}

\section{Introduction}
Blazars are thought to be radio-loud active galactic nuclei (AGNs) which have jets oriented close to the observer line of sight \citep{Urry and Padovani (1995)}.
They emit in all frequencies extending from radio to very high energy $\gamma$-rays.
In general, Blazars shows minutes \citep{Aharonian et al. (2007)} to years 
\citep{Raiteri et al. (2013)} scale of variability time across the entire electromagnetic spectrum. Their spectral energy distribution (SED) is characterized
by two hump kind of structures. The first one peaks in low energy band (IR to soft X-ray), which is well explained by synchrotron emission caused by relativistic
electrons in the magnetic field of the jets and the second one peaks in the high energy band (hard X-ray to $\gamma$-ray), which is thought to be the product
of inverse Compton scattering of low energy photon within the jets called Synchrotron self Compton (\citealt{Konigl (1981)};\citealt{Marscher and Gear (1985)}
;\citealt{Ghisellini and Tavecchio (2009)})
or from outside the jets (External Compton; EC) with relativistic electrons. 
There is also an alternative way that can produce the high energy hump through the hadronic process in which high energy protons interact with low energy
protons and produce charge and neutral muons, and that can decay into high energy $\gamma$-ray and neutrinos. 
In leptonic scenarios, the external seed
photons can come from direct disk emission, BLR, dusty or molecular torus \citep{Bottcher (2007)}. 
\\
Ton 599 is a FSRQ also known as 4C 29.45, and 3FGL J1159.5+2914 (\citealt{Acero et al. (2015)})
with RA = 179.8826413 deg, Dec = 29.2455075 deg, and z = 0.72449. This is the first time the source has gone through a long flaring state across the entire electromagnetic spectrum.
Many correlation studies have been done, between optical, X-ray and $\gamma$-ray to discussed the connection between their emission regions, in FSRQ.
\citet{Cohen et al. (2014)} have studied the correlation between optical and $\gamma$-ray for 40 Blazars and they found that high energy emission leads
the low energy emission with a time lag of 1-10 days. A correlation study of a sample of 183 blazars have also been done by \citet{Pushkarev et al. (2010)} 
and they found that in most of the cases radio flare lags the gamma-ray flare. It is also true that the time delay between flares of two 
bands depends on their separation \citep{Fuhrmann et al. (2014)}. 
In blazar, the exact location of the gamma-ray emission region is not
known because of the poor angular resolution in high energy. While on the other hand, the radio emission region has been resolved in the jets
of blazar with milliarcsecond resolution of radio observations. \citet{Ramakrishnan et al. (2014)} have also been studied the correlation between
$\gamma$-ray and radio emission for this source and they found that $\gamma$-ray is lagging behind the radio with a time lag of 120 days, and that 
constrains the gamma-ray emission region in the parsec-scale jet. \\
In this paper, I have studied the correlation between optical, X-ray and gamma-ray to understand the multi-waveband emission during the flare of 2017.

\section{Multiwavelength Observations and Data Analysis}
\subsection{\textit{Fermi}-LAT}
\textit{Fermi}-LAT is a pair conversion $\gamma$-ray Telescope sensitive to photon energies between 20 MeV to higher than 500 GeV, with a field 
of view of about 2.4 sr (\citealt{Atwood et al. (2009)}). The LAT's field of view covers about 20\% of the sky at any time and it scans the 
whole sky every three hours. The instrument was launched by NASA in 2008 into a near earth orbit. 
Ton 599 was continuously monitored by \textit{Fermi}-LAT since 2008 August. 
The standard data reduction and analysis procedure\footnote{https://fermi.gsfc.nasa.gov/ssc/data/analysis/documentation/} has been followed. 
Other analysis procedure is the same as given in \citet{Prince et al. (2018)}.
I have analyzed the \textit{Fermi}-LAT data from Jan 2014 to Jan 2018 and found that most of the time source was in quiescent state and started
showing major activity at the end of 2017 (Figure 1). At the end of 2015, it shows the flux rising, but that does not last for a long time and 
also the maximum flux was $\sim$4$\times$10$^{-7}$ ph cm$^{-2}$ s$^{-1}$. 
\begin{figure*}
\centering
 \includegraphics[scale=0.28]{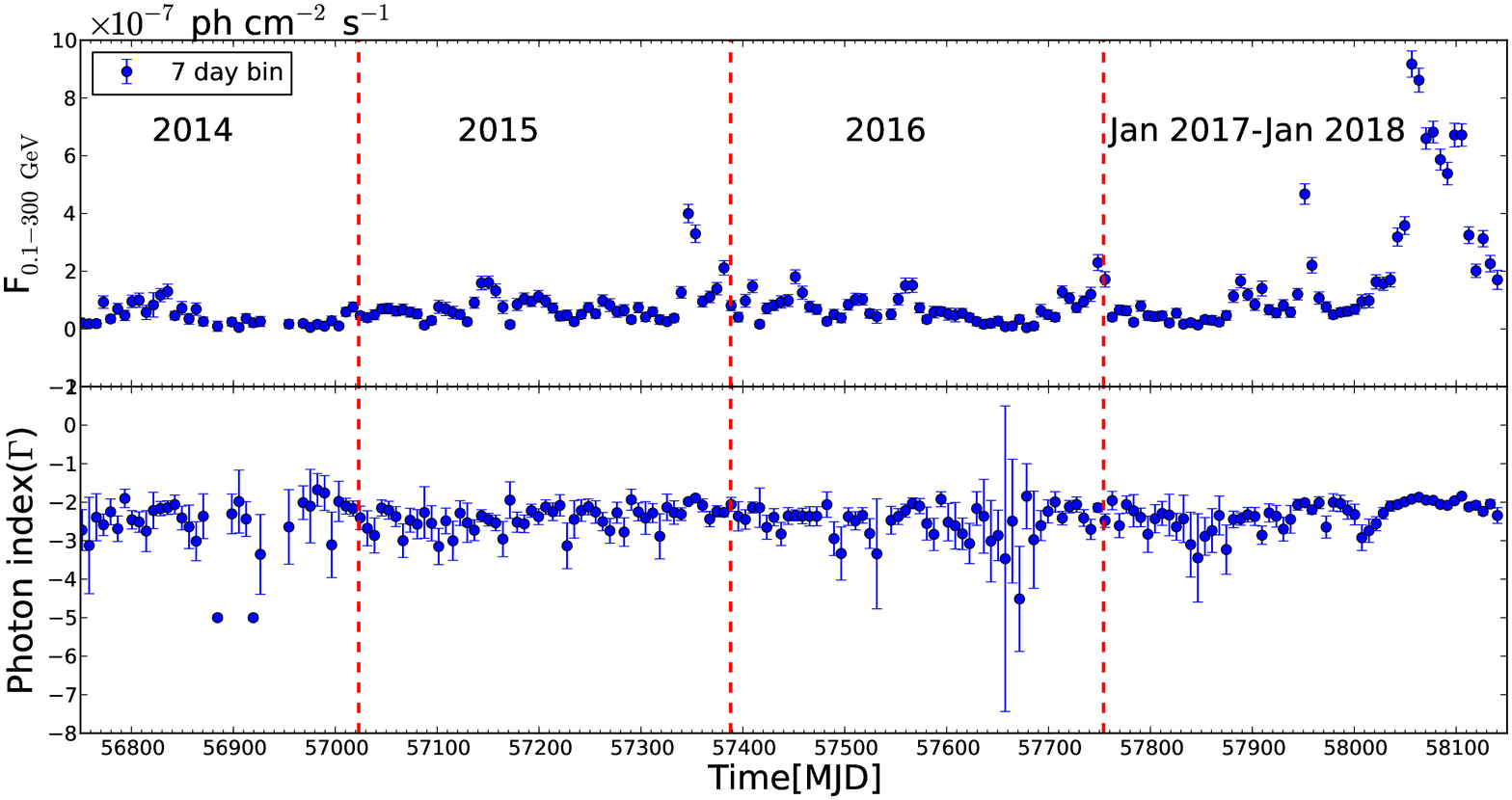}
 \includegraphics[scale=0.28]{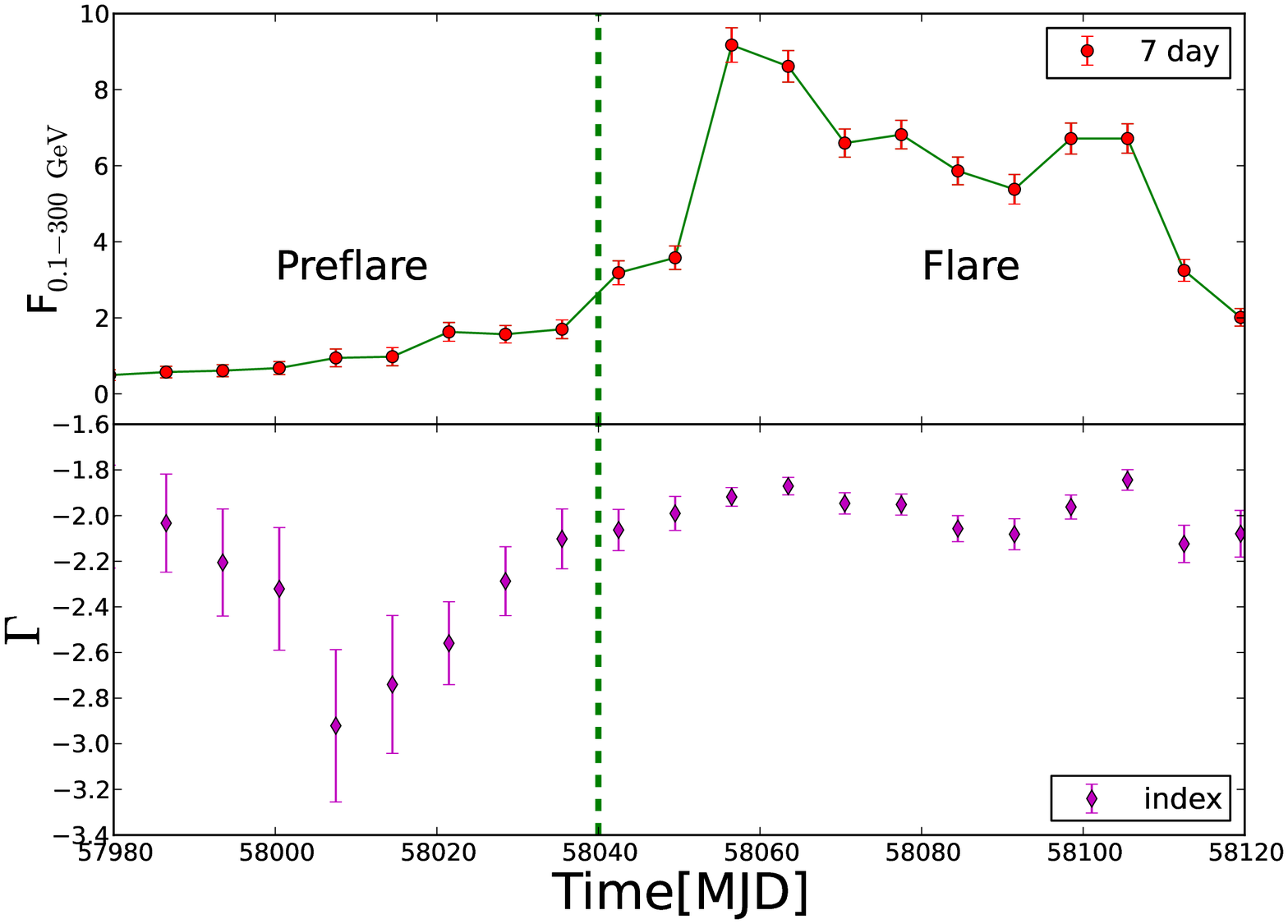}
 \caption{left: Light curve of Ton 599 from Jan 2014 to Jan 2018. right: Zoomed version of flare at the end of 2017 and vertical green dashed 
 line separating the two state of the source. }
\end{figure*}
\begin{figure*}
\centering
 \includegraphics[scale=0.35]{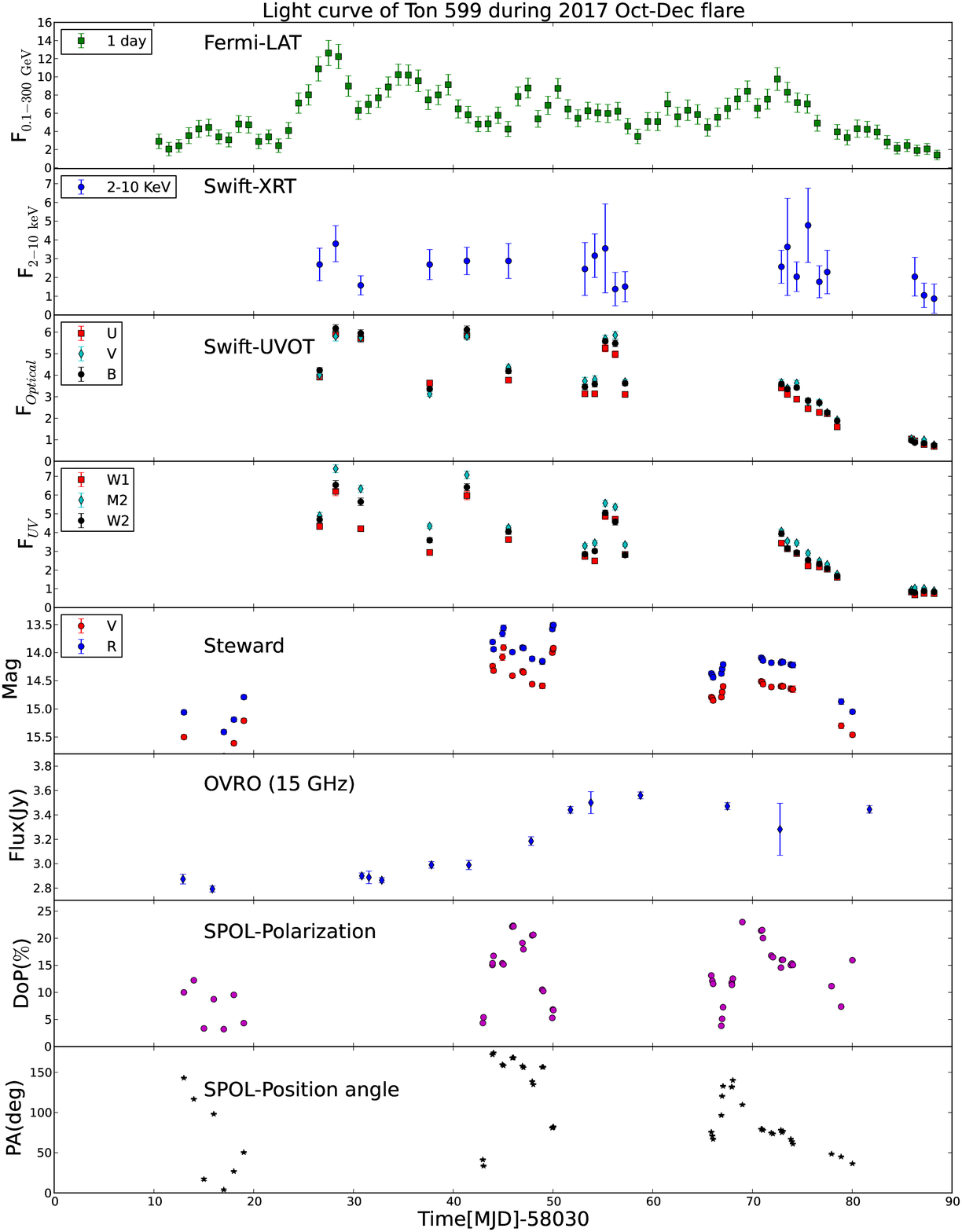}
 \caption{Multi-wavelength light curve of Ton 599 during end of 2017. \textit{Fermi}-LAT data are in units of 10$^{-7}$ ph cm$^{-2}$ s$^{-1}$.
 Swift-XRT and UVOT are in units of 10$^{-12}$ and 10$^{-11}$ erg cm$^{-2}$ s$^{-1}$ respectively.}
 \includegraphics[scale=0.35]{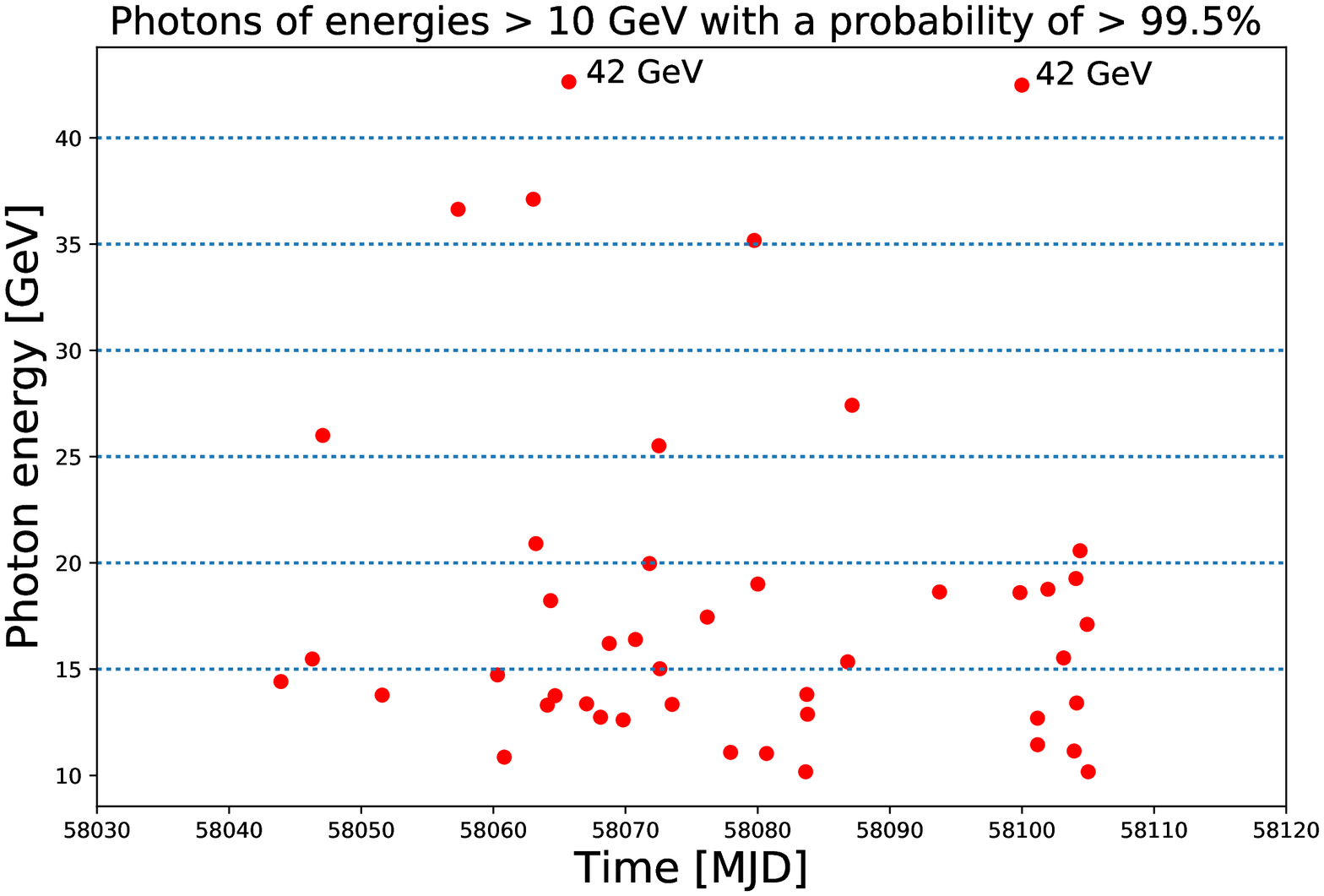}
 \caption{The arrival time of photons of energy $>$ 10 GeV , with probability $>$ 99.5$\%$.}
\end{figure*} 
\subsection{Swift-XRT/UVOT}
Ton 599 was observed by Swift-XRT/UVOT during flaring state. Details of the observations are present in Table 1.
Cleaned event files were obtained using the task `\textit{xrtpipeline}' version 0.13.2.
Latest calibration files (CALDB version 20160609) and standard screening criteria were used for re-processing the raw data. Cleaned event files
corresponding to the Photon Counting (PC) mode were considered. 
Circular regions of radius 20 arc seconds centered at the source and slightly away from the source were 
 chosen for the source and the background regions respectively while analyzing the XRT data.
The X-ray spectra were extracted in \textit{xselect}. 
The obtained spectra is fitted using simple power law model with the galactic absorption column density $n_H$ = 1.77$\times$10$^{20}$ 
cm$^{-2}$ \citep{Kalberla et al. (2005)}. 
The Swift Ultraviolet/Optical Telescope (UVOT, \citealt{Roming et al. (2005)}) also observed Ton 599 in all the six filters U, V, B, W1, M2, and W2. 
The source image was extracted from a region of 5 arc seconds centered at the source. The background region was chosen with a radius of 10 arcseconds
away from the source from a nearby source free region. The `uvotsource'  task has been used to extract the source magnitudes and fluxes. 
Magnitudes are corrected for galactic extinction (\citealt{Schlafly et al. (2011)}) and converted into flux using the zero points 
(\citealt{Breeveld et al. (2011)}) and conversion factors (\citealt{Larionov et al. (2016)}). 

\begin{table}
\centering
\caption{Table shows the log of the observations during the flaring state (MJD 58040 -- 58120).}
 \begin{tabular}{ccc p{1cm}}
 \hline
 && \\
 Observatory & Obs-ID & Exposure (ks)\\
 && \\
 \hline
 && \\
 Swift-XRT/UVOT & 00036381023 & 2.48\\
 Swift-XRT/UVOT & 00036381024 & 2.53\\
 Swift-XRT/UVOT & 00036381025 & 2.46\\
 Swift-XRT/UVOT & 00036381026 & 2.47\\
 Swift-XRT/UVOT & 00036381027 & 2.40\\
 Swift-XRT/UVOT & 00036381028 & 1.61\\
 Swift-XRT/UVOT & 00036381030 & 2.20\\
 Swift-XRT/UVOT & 00036381031 & 2.27\\
 Swift-XRT/UVOT & 00036381032 & 1.58\\
 Swift-XRT/UVOT & 00036381033 & 1.65\\
 Swift-XRT/UVOT & 00036381034 & 1.67\\
 Swift-XRT/UVOT & 00036381035 & 1.99\\
 Swift-XRT/UVOT & 00036381036 & 0.99\\
 Swift-XRT/UVOT & 00036381037 & 1.94\\
 Swift-XRT/UVOT & 00036381038 & 2.02\\
 Swift-XRT/UVOT & 00036381040 & 1.92\\
 Swift-XRT/UVOT & 00036381041 & 0.90\\
 Swift-XRT/UVOT & 00036381042 & 0.87\\
 Swift-XRT/UVOT & 00036381044 & 1.73\\
 Swift-XRT/UVOT & 00036381046 & 1.84\\
 Swift-XRT/UVOT & 00036381047 & 1.94\\
 Swift-XRT/UVOT & 00036381048 & 1.93\\
 &&  \\
 \hline
 
 \end{tabular}
 \label{Table:T1}
\end{table}

\subsection{Steward Optical Observatory}
I have also used the archival data from the Steward optical observatory, Arizona \citep{Smith et al. (2009)}\footnote{http://james.as.arizona.edu/~psmith/Fermi/}. 
Ton 599 is being continuously monitored with the SPOL CCD Imaging/Spectrometer as a part of \textit{Fermi} multi-wavelength support programme.
Optical V-band and R-band photometric data is used along with the Polarimetric (degree of polarization and position angle) data for the whole
flaring period during the end of 2017.

\subsection{OVRO data at 15 GHz}
Ton 599 is also observed in radio by Owens Valley Radio Observatory(OVRO; \citet{Richards et al. (2011)} as a part of \textit{Fermi} monitoring programme.
I have collected the radio data at 15 GHz during MJD 58040 -- 58120. 

\section{Results and Discussions}
I have analyzed the \textit{Fermi}-LAT data from Jan 2014 to Jan 2018 (MJD 56751 -- MJD 58140). The light curve of the source along with the photon index
during these four years is shown in Figure 1. It is clear that most of the time source is in quiescence state where flux is very low (close to zero) and 
sometimes it shows high flux state. Rise in the flux with spectral hardening is observed during the end of 2015 with flux reaching the value 
$\sim$4$\times$10$^{-7}$ ph cm$^{-2}$ s$^{-1}$. After 2015 Ton 599 was in more or less a quiescence state with some small fluctuations in 2016. In 2017
it started showing activity and at the end of 2017 the source under-went clear major flares. 
Zoomed version of the flare is shown in the right panel of Figure 1 and a period has been chosen just before the flare when the source is in
quiescence and called it pre-flare. I have studied this major flare along with
the multi-wavelength observations and done the fractional variability and correlation studies among different wavelength during the flaring
period (MJD 58040 -- MJD 58120). \\
Gamma-ray spectral analysis is also done, four spectral models mentioned in \citet{Prince et al. (2018)} are used to fit the gamma-ray spectral
energy distributions (SEDs).

\subsection{Multi-wavelength light curves}
Multi-wavelength light curve of Ton 599 during the flaring episode MJD 58040 -- MJD 58120 is shown in Figure 2. The first panel shows the 1-day
binning of \textit{Fermi}-LAT data. As we have seen in Figure 1 the source started showing the activity at the end of 2017.
In Figure 2, the flux started rising after MJD 58040 and lasted for around two and a half months and again got back to its quiescence state after MJD 58120.
The flux started rising very slowly and it took around twenty days to become a full-fledged flare. The source showed a clear and major peak at MJD
58057.5 and the maximum flux attained is $\sim$13$\times$10$^{-7}$ ph cm$^{-2}$ s$^{-1}$ from one day binning. After the major peak, the source was in a
higher state for almost two months with an average flux of 6.69$\times$10$^{-7}$ ph cm$^{-2}$ s$^{-1}$.\\
Swift-XRT/UVOT observations were carried out when the source was already in flaring state. 
All the details about the observations are mentioned in Table 1. 
XRT light curve for 2.0-10.0 keV are shown in the second panel of Figure 2. The source shows the higher state in X-ray, and its first peak coincides with 
the $\gamma$-ray peak at MJD 58058 with a flux of 3.80$\times$10$^{-12}$ erg cm$^{-2}$ s$^{-1}$. X-ray flux shows 
fluctuating behavior during the $\gamma$-ray flare and it settled down in its quiescence state as the $\gamma$-ray flare ended.
The quiescence state flux is noted as 0.87$\times$10$^{-12}$ erg cm$^{-2}$ s$^{-1}$.\\
Ton 599 was also observed with UVOT in all the six filters (U, B, V, W1, M2, W2). The light curves for optical (U, B, V) and UV (W1, M2, W2) 
filters are shown in the third and fourth panel of Figure 2 respectively. Since Ton 599 was already flaring when Swift started looking at it, 
so the optical and UV fluxes were already in the high state. It shows the peak at MJD 58058 which clearly coincides with the X-ray as well as 
$\gamma$-ray first peak. At the peak optical U, B, V fluxes are 5.92$\times$10$^{-11}$, 6.17$\times$10$^{-11}$, 5.81$\times$10$^{-11}$ 
and UV W1, M2, W2 fluxes are 6.19$\times$10$^{-11}$, 7.41$\times$10$^{-11}$, 6.54$\times$10$^{-11}$ erg cm$^{-2}$ s$^{-1}$ respectively. 
In the light curve, it is clearly seen that the source was very variable in both optical and UV throughout the whole flaring period similar to 
the $\gamma$-ray. Optical and UV also follow the last peak of $\gamma$-ray flare at MJD 58103.
After two months of flaring period optical and UV, flux attained its quiescence state with a flux close to zero at MJD 58118.\\
Steward V and R band magnitudes are plotted in the fifth panel of Figure 2. It is found that Ton 599 is more bright in R band than V band during 
the flare. The average magnitude during the flare in V and R band are 14.9 and 14.5 respectively. \\
In the sixth panel of Figure 2, the radio light curve is shown at 15 GHz from Owens Valley Radio Observatory (OVRO). It has been clearly seen 
that the source is in quiescence state in radio while it is flaring in the $\gamma$-ray and other waveband. The radio flux started rising slowly 
at MJD 58060 and after almost thirty days it attained the maximum flux 3.56 Jy at MJD 58089. The delay in the radio flare is investigated while 
studying the correlations among the different wave band in section 3.6. \\
The Degree of polarization (DoP) and position angle (PA) are plotted in panel seven \& eight of Figure 2. Huge variation is seen in DoP and PA during the
flare. In 10 days of span MJD 58070--58080, DoP varies from 4\%-- 22\% and PA varies from 30 degree--175 degree.
The variation in the DoP and PA can be explained by shock-in-jet model (\citealt{Marscher et al. (2008)}; \citealt{Larionov et al. (2013)}; \citealt{Casadio et al. (2015)}). 
In which a shock wave moving down the jet following magnetic field lines, covering only a portion of the jet's cross section can lead to this variation 
in DoP and PA during the flare.

\begin{figure}
\centering
 \includegraphics[scale=.24]{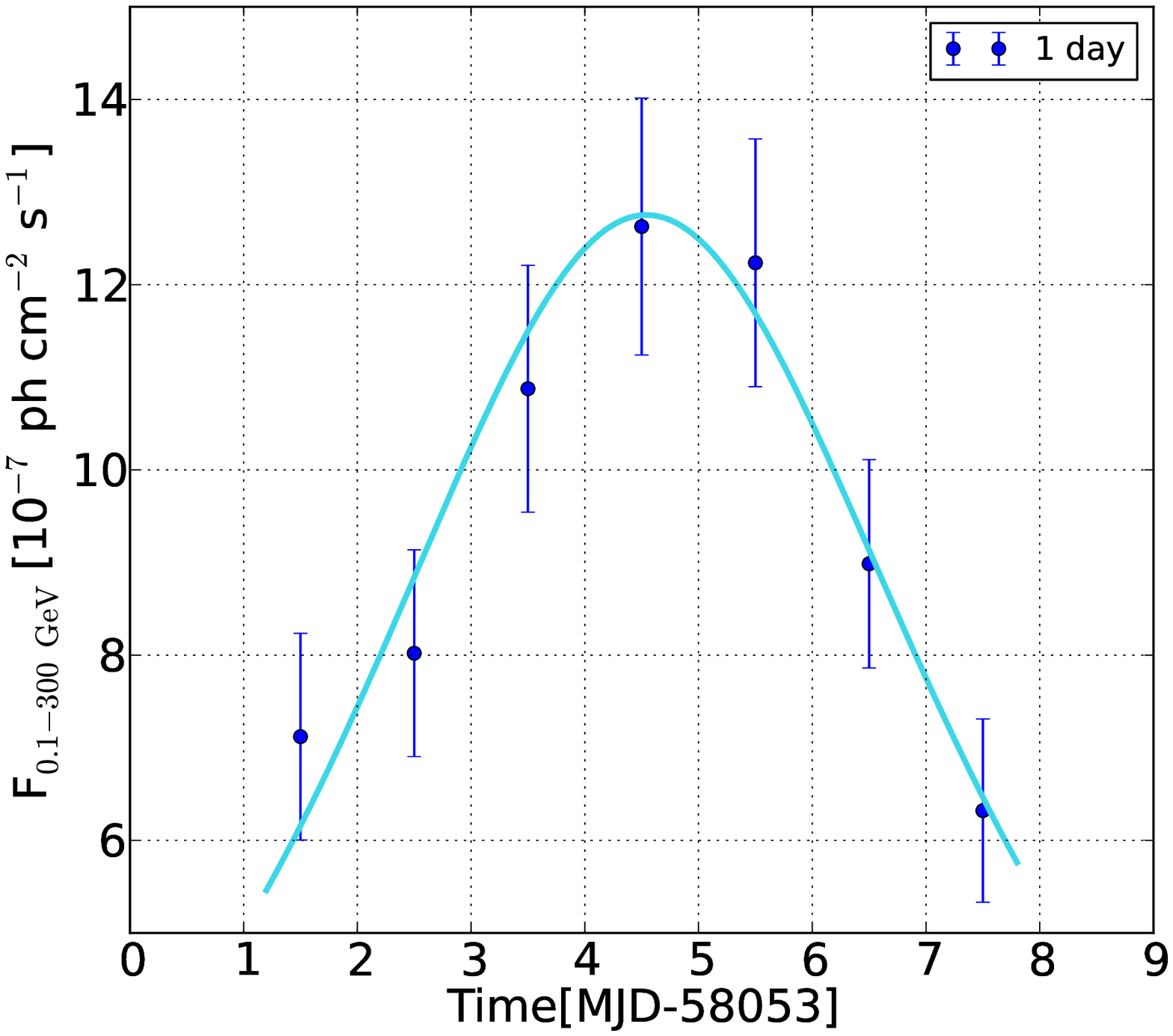}
 \includegraphics[scale=0.24]{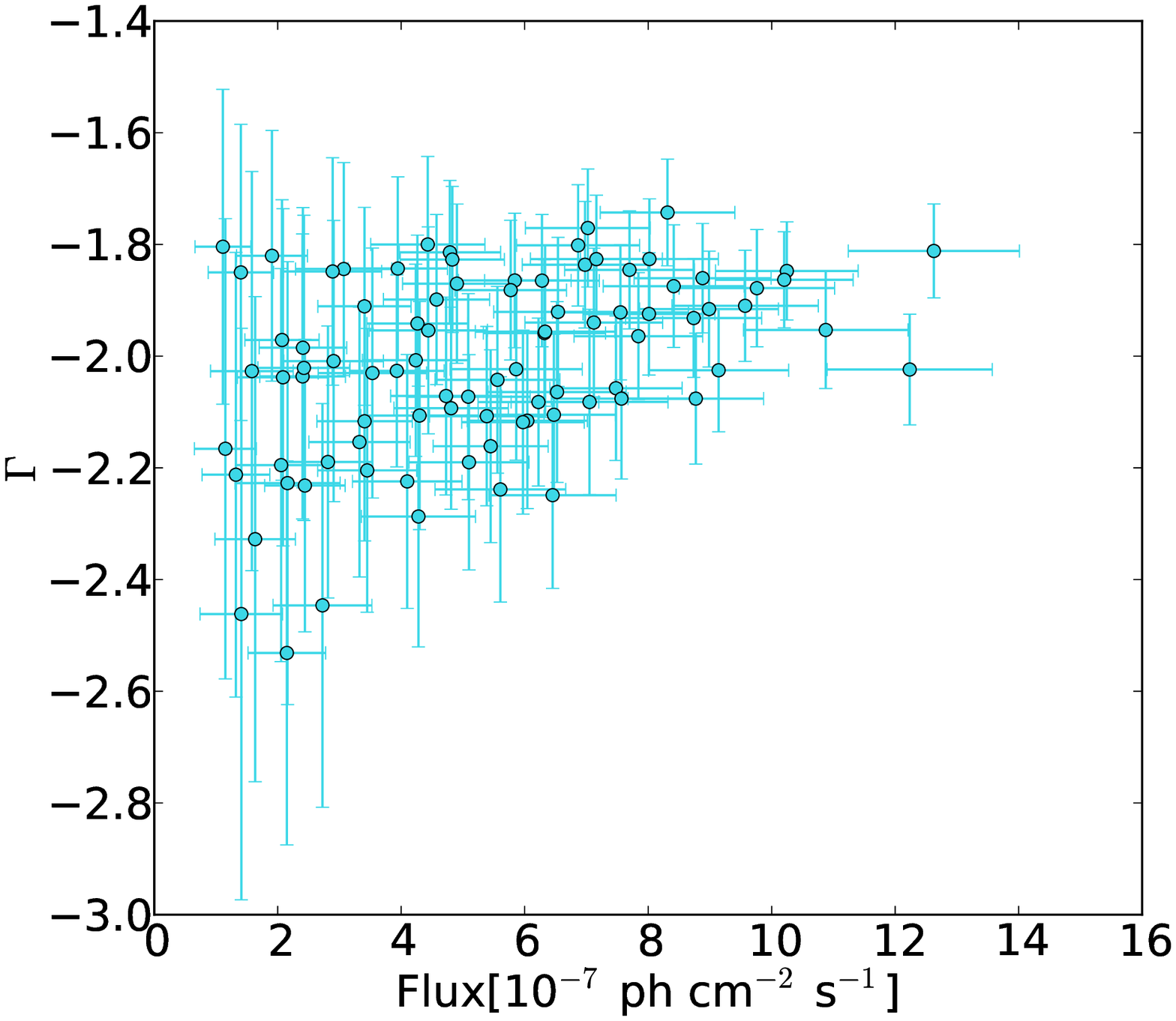}
 \caption{left: Temporal fitting of the flare's first peak. Right: Photon index is plotted with photon flux to show the brighter and harder trend.  }
\end{figure}

\subsection{High energy photons and temporal evolution}
High energy photons are also detected by \textit{Fermi}, using the ''ULTRACLEAN'' class of events and 0.5$^{0}$ of ROI. The results are plotted in Figure 3, 
that shows the photons energy on y-axis and their arrival time on the x-axis. Photons of energy greater than 10 GeV
and with a probability of greater than 99.5$\%$ are only shown in Figure 3. It is found that most of the photons have energy below 20 GeV and only
few have been detected above 20 GeV. Two 42 GeV of photons have been detected during the flare with a probability of 99.7$\%$ and 99.8$\%$ at MJD
58065.7 and 58100 respectively.
\\
The temporal evolution of flare has been studied here and, I have fitted the first peak of the flare, shown in Figure 4,
by a sum of exponentials which provides the rise and decay time of the peak. The functional form of sum of exponentials is as follows:

\begin{equation} \label{1}
 F(t) = 2F_0 \left[\rm{exp}(\frac{t_0-t}{T_r}) + exp(\frac{t-t_0}{T_d})\right]^{-1},
\end{equation}
 where $F_0$ = flare amplitude at time $t_0$, $T_r$ = rise time, $T_d$ = decay time \citep{Abdo et al. (2010)}.
Peak shown in Figure 4 is symmetric with rising and decay time of 2.22$\pm$0.14 and 2.30$\pm$0.13 days respectively. 
The temporal fitting is also applied for other peaks found during the flare, and most of them are found to be symmetric.
The symmetric time profile is expected when the cooling time of electrons t$_{cool}$ is much smaller than the light crossing time R/c 
\citep{Chiaberge and Ghisellini (1999)}, where R is the size of the emission region. \\
In the lower panel of Figure 4, Gamma-ray fluxes are plotted with respect to the photon spectral index and a clear brighter and harder spectral 
behavior is seen. During this high activity, the spectral index is harder than those reported in 3FGL catalog \citet{Acero et al. (2015)} for 
this source. 

\subsection{$\gamma$-ray emission region}
The $\gamma$-ray flare along with photon spectral index is plotted separately in Figure 5. A clear variation in the spectral index is seen during the
period (MJD 58040 -- MJD 58120).
In the pre-flare state (Figure 1), between MJD 57980 to MJD 58040, the source is almost in quiescence with an average flux of 
9.35$\times$10$^{-8}$ ph cm$^{-2}$ s$^{-1}$ and the average photon index is 2.38. Once the source is in full fledge flaring episode 58054 to 58110. 
The average photon flux rises to 6.94$\times$10$^{-7}$ ph cm$^{-2}$ s$^{-1}$ with an average spectral index of 1.96. 
The maximum flux attained during flaring episode is 12.63$\times$10$^{-7}$ ph cm$^{-2}$ s$^{-1}$ at MJD 58057.5 with photon index 1.81 (Figure 5).
The fastest variability time from the 1-day binning light curve (Figure 5) is estimated here, by using the following expression
\begin{equation} \label{2}
F(t_2) = F(t_1).2^{(t_2-t_1)/ t_{d}},\\
\end{equation}
where $F(t_1)$ and $F(t_2)$ are the fluxes measured at time $t_1$ and $t_2$ respectively and $t_d$ represents 
the doubling/halving timescale of flux. A range of variability time is found, from one day to a few days. One day is used as the fastest variability
time to estimate the size of the emission region, by using the relation 
\begin{equation}
 R \leq c\, t_{var} \, \delta \, (1+z)^{-1}
 \end{equation}
where, z = 0.72 and $\delta$ is the Doppler factor. The size of the emission region is found to be 1.88$\times$10$^{16}$ cm, for
$\delta$ = 12.5 (\citealt{Zhang et al. (2002)}; \citealt{Liodakis et al. (2017)}) which is close to value ($\delta$ = 15) estimated by \citet{Ghisellini et al. (1998)}.

Detection of high energy photons ($>$ 20 GeV) during the flare of Ton 599 puts a constraint on the location of the $\gamma$-ray emission region.
\citet{Liu and Bai (2006)} have estimated the optical depth for gamma-rays with energies 10-100 GeV produced within the BLR. They have found that
the BLR is opaque for above 20GeV/(1+z) gamma-ray photons. This means that the high energy photons seen during the flare (Figure 3) must have been 
emitted outside or outer edge of the BLR.
Distance (R) of the $\gamma$-ray emitting blob from the central supermassive black hole is also estimated by using the relation R $\sim$ r/$\psi$,
where 'r' is the size of the $\gamma$-ray emitting region and $\psi$ is the semi-aperture angle of the jet (\citealt{Foschini et al. (2011)}).
In general, $\psi$ lies between 0.10-0.25 (\citealt{Ghisellini and Tavecchio (2009)}; \citealt{Dermer et al. (2009)}).

 
The intrinsic opening angle is estimated from observations by \citet{Pushkarev et al. (2009)}, and they found that the average intrinsic
opening angles for a sample of BL Lacs is 2.4$\pm$0.6 degree and for quasars is 1.2$\pm$0.1 degree. In \citet{Pushkarev et al. (2009)} sample, 
Ton 599 is listed as J1156+295, and the intrinsic opening angle is derived as 0.58$\degree$. For opening angle 0.58$\degree$, the location of emission
region is estimated as 3.24$\times$10$^{16}$ cm which is near the boundary of the BLR (2.4$\times$10$^{17}$ and 2.98$\times$10$^{17}$ cm) dissipation region 
estimated by \citet{Wu et al. (2018)} and \citet{Pian et al. (2005)} respectively.
Therefore, at the time of 42 GeV of photon emission during the flare, $\gamma$-ray emission region must have been located outside
or at the edge of the BLR.
\citet{Pushkarev et al. (2017)} have also calculated
the intrinsic opening angle for 65 sources from MOJAVE-1 sample. They have found that the intrinsic opening angles for these 65 sources lie between
0.1$\degree$ to 9.4$\degree$, with a median of 1.3$\degree$. The range of opening angle suggest the location of the emission region must lie between 
1.88$\pm$10$^{17}$ to 2.00$\pm$10$^{15}$ cm. The location of the emission region estimated for Ton 599 (3.24$\times$10$^{16}$ cm) 
is found to be in this range.

\begin{figure}
 \includegraphics[scale=0.31]{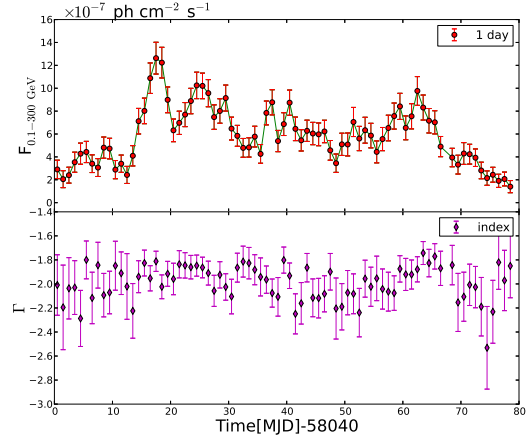}
 \caption{One day bin gamma-ray LC with photon spectral index.}
 \end{figure} 
 
\begin{figure}
\centering
  \includegraphics[scale=0.3]{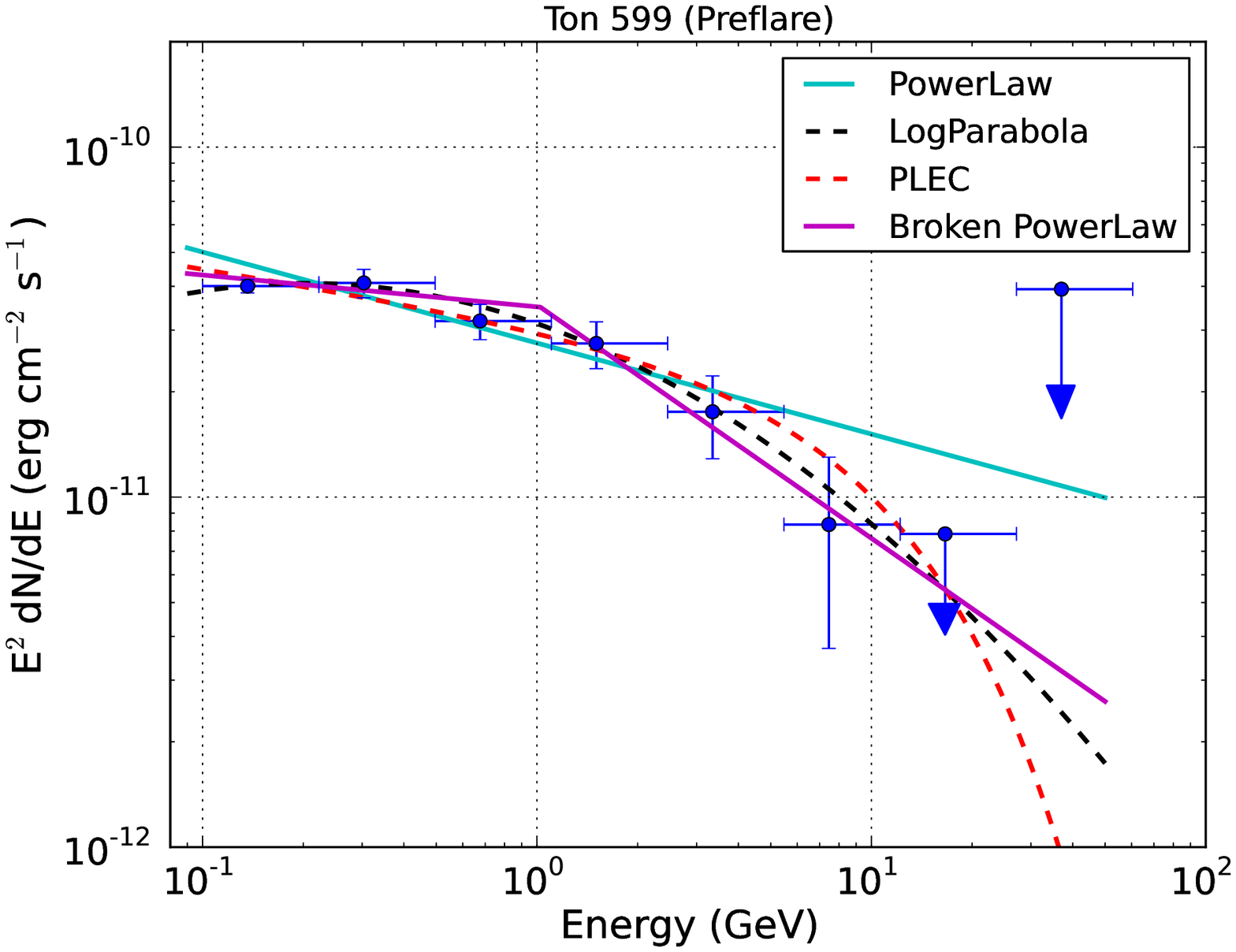}
 \includegraphics[scale=0.3]{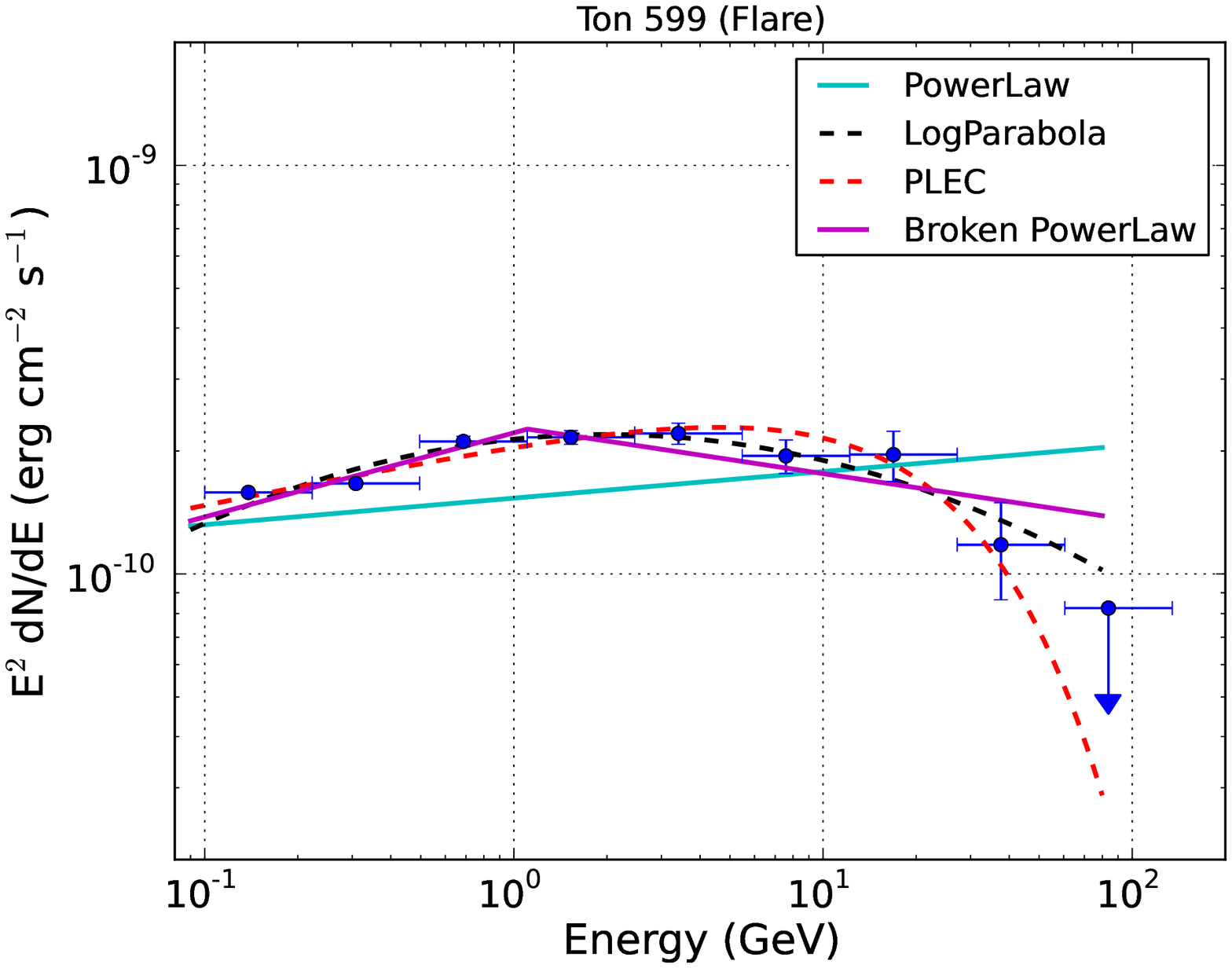}
 \caption{SEDs of pre-flare and flaring period of Ton 599}
\end{figure} 

\begin{table*}
 \caption{Parameters obtained from the spectral analysis fit, for the different models PL, LP, PLEC, and BPL, for the pre-flare and flare by using 
 the Likelihood analysis method. $\Delta$Log(likelihood) is estimated with respect to the Log(likelihood) of the PL fit.}
 \begin{tabular}{c c c c c c c}
 \hline 
  &&PowerLaw (PL)&&&&  \\
\\
Activity  &  F$_{0.1-300}$ $_{\rm{GeV}} $ & $\Gamma$ & && -Log(likelihood) & $\Delta$Log(likelihood) \\[0.5ex]
\\
& ($10^{-7}$ ph cm$^{-2}$ s$^{-1}$)&&&&  \\

\hline
\\
Pre-flare& 2.45$\pm$0.15 & 2.26$\pm$0.05  & && 99810.26&  \\
\\
Flare& 11.00$\pm$0.02 &1.94$\pm$0.01 & & &183323.43 &   \\
\\
\hline
&&LogParabola (LP)&&&  \\
\\
  &   & $\alpha$ & $\beta$& &   &  \\[0.5ex]
\hline
\\
Pre-flare&2.27$\pm$0.01 & 2.08$\pm$0.09  & 0.10$\pm$0.04 && 99806.43& -3.83  \\
\\
Flare& 10.40$\pm$0.02 & 1.79$\pm$0.02 &0.06$\pm$0.01 & &183294.99 & -28.44  \\
\\
\hline
&&PLExpCutoff (PLEC)&&&  \\
\\
&   &$\Gamma_{PLEC}$ & E$_{cutoff}$ && &   \\
\hline
\\
Pre-flare& 2.36$\pm$0.02 & 2.15$\pm$0.08 &  12.51$\pm$7.60 && 99808.07 & -2.19 \\
\\
Flare& 10.60$\pm$0.02 &1.85$\pm$0.01 &30.00$\pm$0.08 & &183291.44 & -31.99  \\
\\
\hline
&&Broken PowerLaw (BPL)&&&  \\
\\
&   &$\Gamma_{1}$ & $\Gamma_{2}$ & E$_{break}$ & &   \\
\hline
\\
Pre-flare& 2.30$\pm$0.02 & 2.09$\pm$0.08 & 2.67$\pm$0.17 & 1.10$\pm$0.17  & 99806.06 & -4.2  \\
\\
Flare& 10.50$\pm$0.02 &1.79$\pm$0.03 &2.11$\pm$0.04 &1.11$\pm$0.22 &183297.07 & -26.36   \\
\\
\hline
\\

\end{tabular}

\end{table*}

\subsection{Spectral Analysis}
The spectral analysis of pre-flare and flare observed at the end of 2017 is presented in this section. Likelihood analysis is done with four different
spectral models mentioned in \citet{Prince et al. (2018)}. The SEDs data points are fitted with four spectral models (PL, LP, PLEC, and BPL)
discussed in \citet{Prince et al. (2018)},
the fitted parameters are presented in Table 2, and the plots are shown in Figure 6.
Fitting the gamma-ray SEDs data points with these four models will help us to constrains the gamma-ray emission region. Inside the BLR,
photon-photon pair production ($\gamma$ $\gamma$ $\longrightarrow$ e$^{+}$e$^{-}$) can attenuate the gamma-ray flux and as a result, we expect to see
a break in gamma-ray spectrum. A break in the gamma-ray spectrum can be examined by fitting the gamma-ray SEDs data points with LP/BPL/PEC. While in other 
case when the emission region is outside the BLR or within the molecular torus (MT) a simple PL could be a good fit to the SED data points.
In Table 2, the quality of the unbinned fit is presented by the 
Log(likelihood) value and the model with a large value of $\Delta$Log(likelihood), with respect to PL, preferred over the lower one. Overall all 
the three models LP, PLEC, and BPL are compatible with the SEDs data points. A clear spectral hardening is seen with increasing flux when the source
travels from pre-flare to flaring state. For PL, during pre-flare to flare the flux rises from 2.45$\pm$0.15 to 11.00$\pm$0.02 ($\times$10$^{-7}$ 
ph cm$^{-2}$ s$^{-1}$) and the spectral index ($\Gamma$) changes from 2.26$\pm$0.05 to 1.94$\pm$0.01.
A break in the $\gamma$-ray spectrum is seen during the flare while fitting the SED with BPL. It shows the rising spectrum before the break and falling
spectrum after the break. Before the break, the BPL photon index $\Gamma_1$ is 1.79, the break energy E$_{break}$ is 1.11 GeV, and after the break, 
BPL photon index $\Gamma_2$ is 2.11. This suggests that the peak of the IC mechanism
probably lie in the LAT energy band and the shape of the $\gamma$-ray spectrum likely reflects the distribution of emitting electrons.

\subsection{Fractional variability (F$_{var}$)}
Variability seen at all frequencies and timescales in blazars is completely a random process. It is more prominent during the flare, and the flare
profiles depend on the particle acceleration and energy dissipation. The amplitude of variation depends on the jet parameters like magnetic fields,
viewing angle, particle density and the efficiency of acceleration \citep{Kaur and Baliyan (2018)}. To determine the variability amplitude in all 
energy band, good quality data is required across the entire electromagnetic spectrum. Observation of Ton 599 across the entire electromagnetic
spectrum makes it possible to determine the variability amplitude using the fractional root mean square (rms) variability parameter (F$_{var}$) introduced
by \citet{Edelson and Malkan (1987)}; \citet{Edelson et al. (1990)}.

Fractional variability is used to compare the variability amplitudes across the entire electromagnetic spectrum 
and can be estimated by using the relation given in \citet{Vaughan et al.(2003)},
\begin{equation} \label{8}
 F_{var} = \sqrt{\frac{S^2 - \sigma^2}{r^2}} \\
\end{equation}

\begin{equation}
 err(F_{var}) = \sqrt{  \Big(\sqrt{\frac{1}{2N}}. \frac{\sigma^2}{r^2F_{var}} \Big)^2 + \Big( \sqrt{\frac{\sigma^2}{N}}. \frac{1}{r} \Big)^2     } \\
\end{equation}

where, $\sigma^2_{XS}$ = S$^{2}$ -- $\sigma^2$, is called excess variance, S$^{2}$ is the sample variance, $\sigma^2$ is the mean square uncertainties 
of each observations and r is the sample mean. 

\begin{table}
\centering
\caption{Fractional variability is estimated for time interval 57980 to 58120.}
\begin{tabular}{c c c c c}
 \hline
 \hline
 \\
 Waveband  & F$_{var}$ & err(F$_{var}$) \\
 \\
 \hline
 
 $\gamma$-ray &  0.730 & 0.019 \\ 

 U & 0.514& 0.008 \\
 
 B & 0.503& 0.007 \\

 V & 0.485& 0.008 \\
 
 W1 & 0.537& 0.009 \\
 
 M2 & 0.531& 0.007 \\
 
 W2 & 0.536& 0.008 \\
 
 OVRO (15 GHz) & 0.071 & 0.004 \\
 
 \hline
\end{tabular}
\label{Table:TA}
\end{table}

The fractional variability for all the wavebands is mentioned in Table 3.
It is clear that source is most variable in $\gamma$-ray and then UV, Optical and radio (at 15 GHz). 
Because of the large error bar in the X-ray data, I could not estimate the fractional variability.
The F$_{var}$ is 0.73 in $\gamma$-ray, 0.53 in UVW2-band, 0.50 in optical B-band and 0.07 in radio (at 15 GHz). 
It is found that F$_{var}$ is increasing with energy, suggesting that a large number of particles are producing 
the high energy emission. Similar behavior of fractional variability is also seen for other FSRQ like CTA 102 by 
\citet{Kaur and Baliyan (2018)}, where they found a trend of large fractional variability towards higher energies. Increase in fractional 
variability is also seen in TeV blazar, 
\citet{Patel et al. (2018)} and \citet{Sinha et al. (2016)} have noted an increase in fractional variability from radio to X-rays and decrease 
in high energy part from $\gamma$-rays to Hard X-rays. An opposite trend was also reported by \citet{Bonning et al. (2009)}, where variability 
amplitudes decrease towards shorter wavelength (IR, Optical, and UV), which suggests the presence of steady thermal emission from the accretion disk.

\subsection{Correlations}
From Figure 2, it is very clear that the flares in $\gamma$-ray, X-ray, Optical and UV band are mostly correlated. The radio flare at 15 GHz 
noted after few days of $\gamma$-ray flare. The detailed study about correlations has been done in this particular section.
A cross-correlation study of flux variations in different energy band can give an idea of whether emissions in different bands are coming from
the same emission region in the jet and if not then it gives an indication of a relative distance between the emitting zones.
So, I have done the correlation studies using the discrete correlations function (zDCF) formulated by 
\citet{Edelson and Krolik (1988)}. It provides insight about the emission in different energy band.
Let's suppose there are two discrete data sets a$_i$ and b$_j$ and they have standard deviation $\sigma_a$ and $\sigma_b$, the discrete
correlations for all measured pairs (a$_i$-b$_j$) is defined as,
\begin{equation}
 UDCF_{ij} = \frac{(a_i-\bar{a})(b_j-\bar{b})}{\sqrt{(\sigma_a^2-e_a^2)(\sigma_b^2-e_b^2)}}
\end{equation}
Where each pairs are associated with a pairwise lag $\Delta t_{ij}$ = t$_j$ - t$_i$. The parameters e$_a$ and e$_b$ are the measurement errors 
associated with data sets a$_i$ and b$_j$ respectively. Binning the UDCF$_{ij}$ in time will directly result in DCF($\tau$). Averaging the UDCF$_{ij}$ over M
number of pairs for which ($\tau$ - $\Delta \tau$/2) $\leq$ $\Delta t_{ij}$ $<$ ($\tau$ + $\Delta \tau$/2),
\begin{equation}
 DCF(\tau) = \frac{1}{M} UDCF_{ij},
\end{equation}
and the error on DCF is defined as,
\begin{equation}
 \sigma_{DCF}(\tau) = \frac{1}{M-1} \Bigg\{\sum[UDCF_{ij} - DCF(\tau)] \Bigg\}^{1/2}
\end{equation}

\begin{figure*}
\centering
 \includegraphics[scale=0.30]{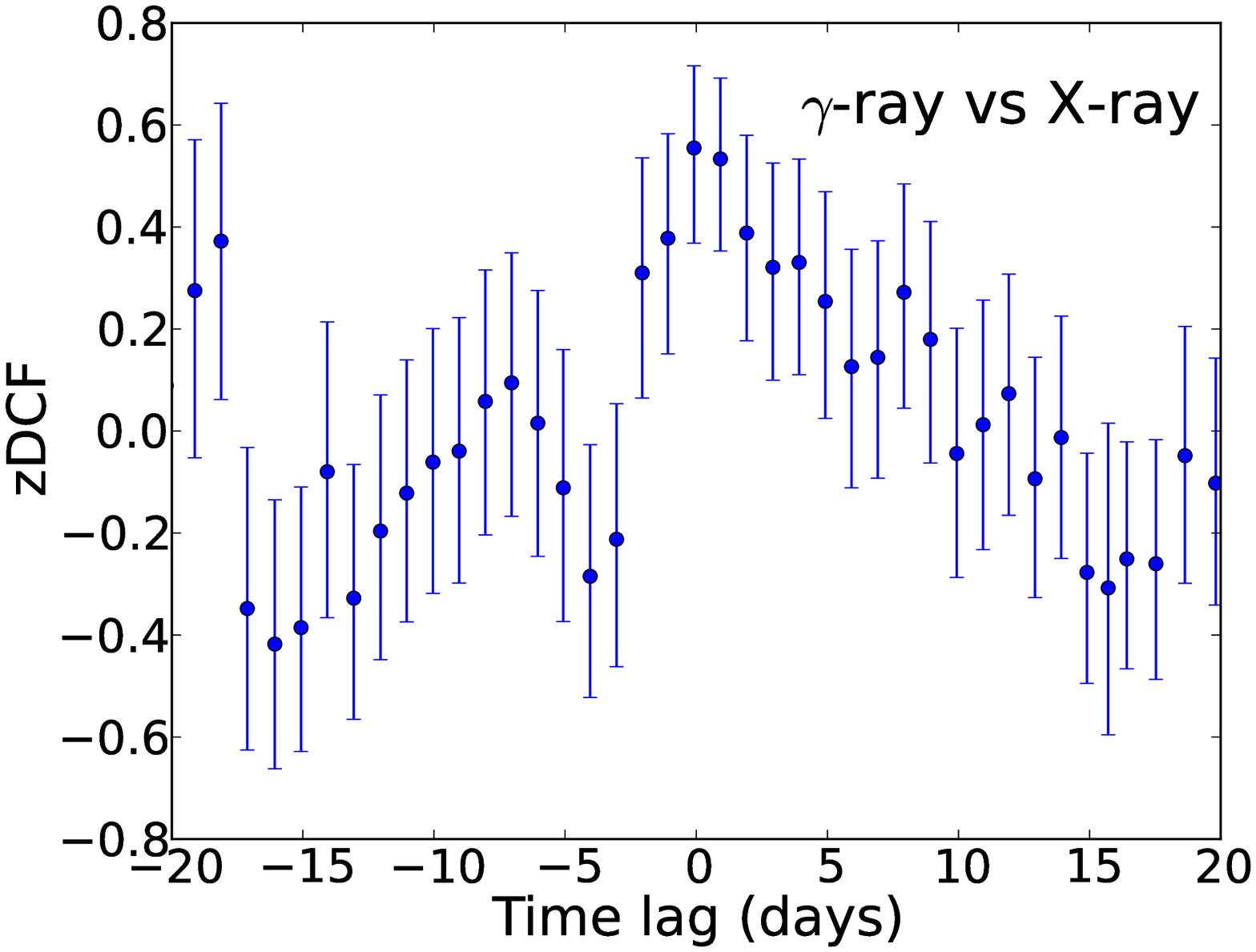}
 \includegraphics[scale=0.30]{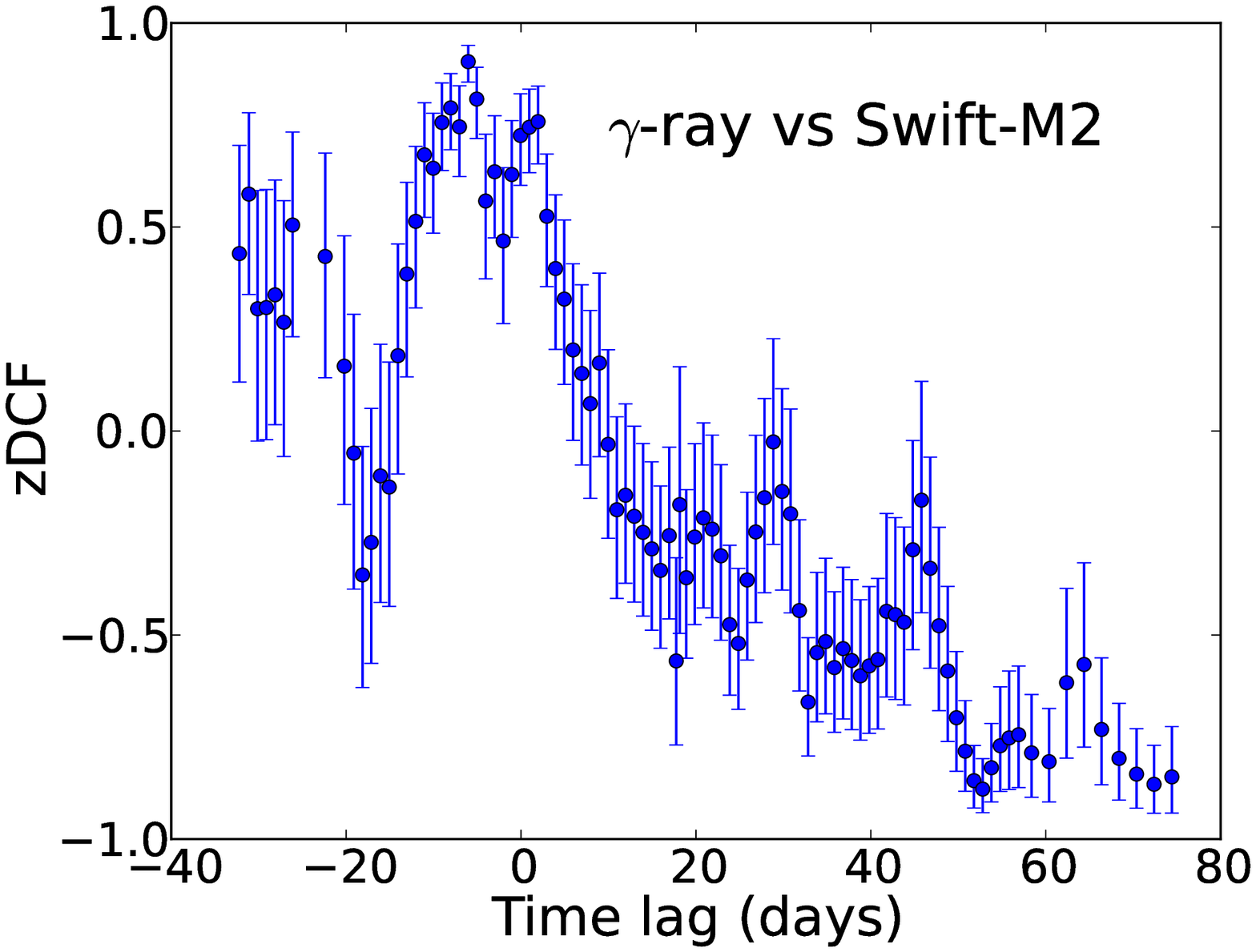} \\
 \includegraphics[scale=0.30]{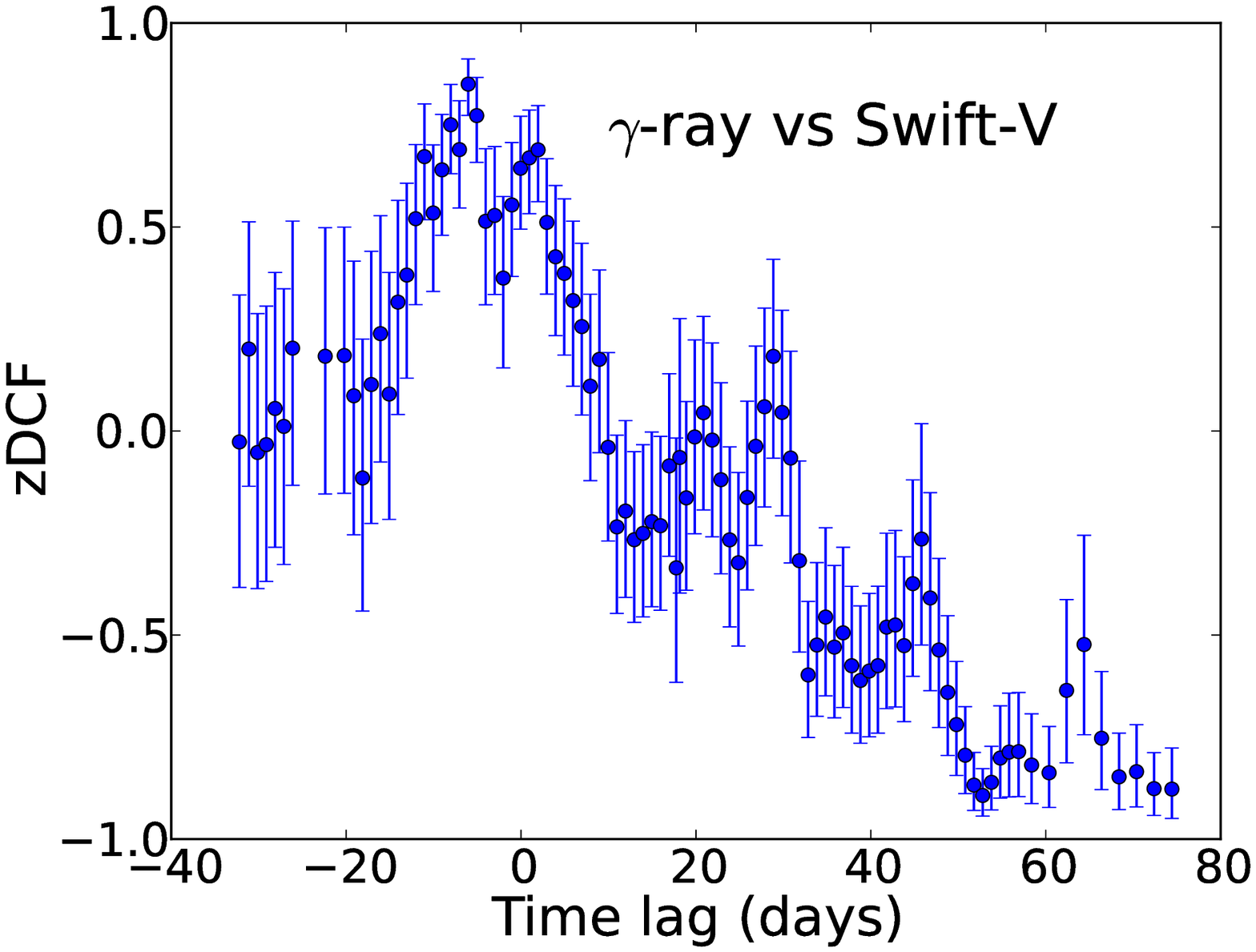}
 \includegraphics[scale=0.30]{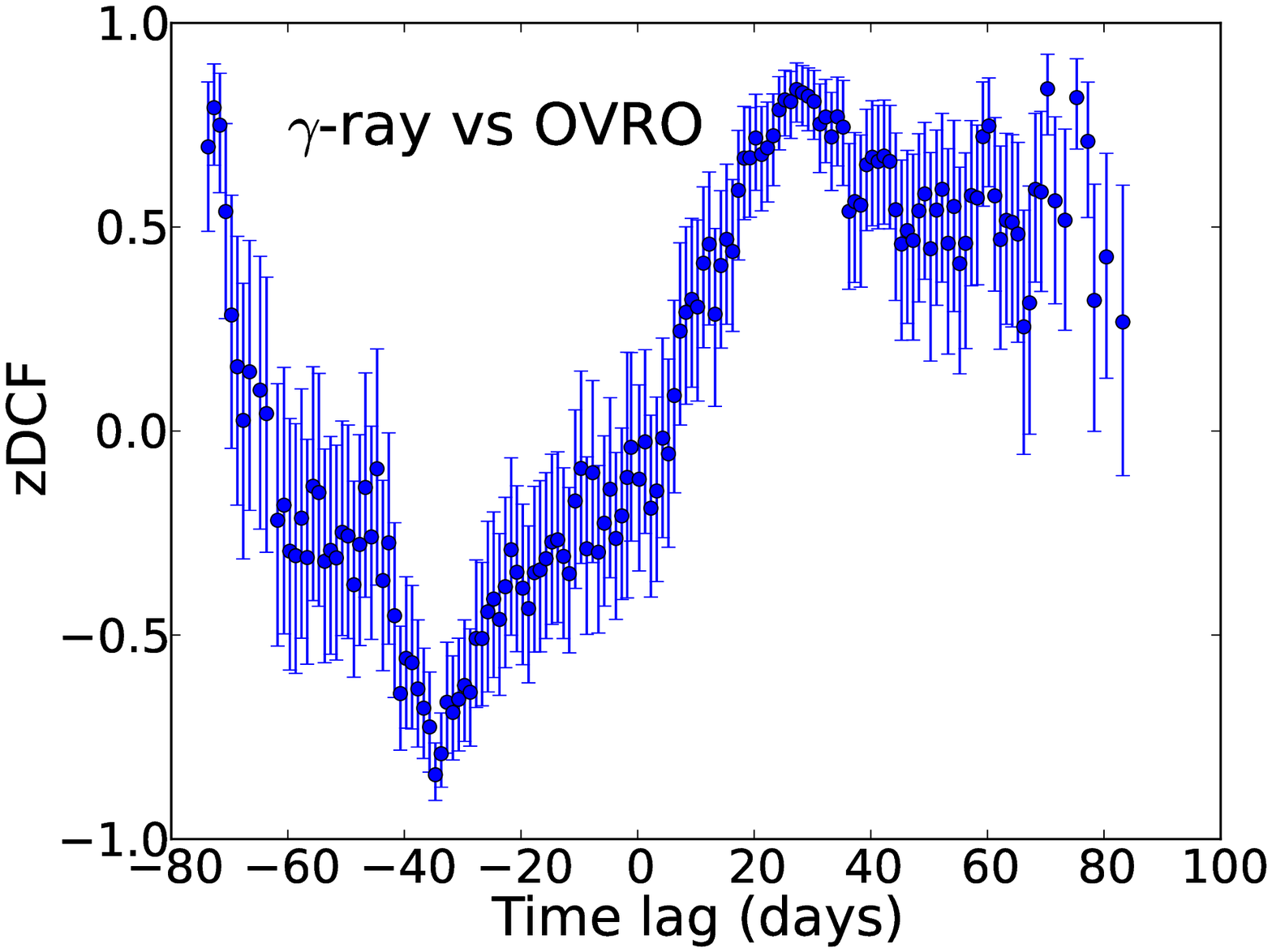}
 \caption{DCF is plotted for all four combinations: $\gamma$-X-ray, $\gamma$-Swift-M2, $\gamma$-Swift-V, $\gamma$-OVRO(15 GHz) from left to right,
 for the flare of Ton 599 during end of 2017.}
\end{figure*}

Discrete correlations function (DCF) are plotted in Figure 7 for different combinations like $\gamma$-X-ray, $\gamma$-Swift M2, $\gamma$-Swift V and
$\gamma$-OVRO (15 GHz).\\ In $\gamma$-X-ray correlations it is found that there is no time lag between $\gamma$-ray and X-ray emission and the maximum DCF 
is 0.55. The strong correlation and zero time lag between $\gamma$-ray and X-ray suggests that the emissions are originated from the same region or very close-by region.\\
A Significant correlation has been seen in $\gamma$-ray and optical (V-band) emission with a small time lag and the peak DCF is noted as 0.85. Similar
kind of behavior is also seen in $\gamma$-ray and UV (M2 filter) emission with peak DCF 0.90. \citet{Larionov et al. (2013)} also found small lag in
$\gamma$-ray and optical emission for S50716+71 and at the same time they also noted an emergence of radio knot K3. Finally, they have concluded that all 
these events are co-spatial. Similar results were also noticed for CTA 102 (\citealt{Larionov et al. (2016)}; \citealt{Kaur and Baliyan (2018)}) during the 
outburst of 2012 and 2017 with remarkable similarity in two energy emission. Significant correlation and small time lag in $\gamma$-ray and optical/UV can be explained 
by leptonic models, where it is assumed that the optical/UV emission is mostly the synchrotron emission from the jets and the $\gamma$-ray 
emission is the product of inverse Compton (IC) scattering of optical/UV photons by the relativistic electrons present in the jets.  
\\
It is believed that $\gamma$-ray emission is product of IC scattering of soft photons off the same electrons producing the optical
radiation, then its variations are expected to be simultaneous or delayed with respect to the optical radiation, and it can be the result 
of modeling the non-thermal flares with shocks in a jet model (\citealt{Sikora et al. (2001)}; \citealt{Sokolov et al. (2004)}; \citealt{Sokolov and Marscher (2005)}). 
 This kind of behaviour is already seen in few other blazars like 4C 38.42 (\citealt{Raiteri et al. (2012)}), 3C 345 (\citealt{Schinzel et al. (2012)})
 and in 3C 454.3 (\citealt{Bonning et al. (2009)}; \citealt{Vercellone et al. (2010)}; \citealt{Raiteri et al. (2011)}). Interestingly, the 
 opposite behaviour is also seen, where $\gamma$-ray is leading with optical
 radiation, in few blazars e.g. FSRQ PKS 1510-089 (\citealt{Abdo et al. (2010a)}; \citealt{D'Ammando et al. (2011)}) and in 3C 279 \citep{Hayasida et al. (2012)}. 
 It can be explained by considering fast decay in energy density of external seed photon, responsible for the IC emission, along with the jet 
 axis, compared to the decay in magnetic field energy density which is responsible for the synchrotron emission.\\
 A complex correlation between gamma-ray and optical radiation has also been addressed, by \citet{Marscher (2014)}, by considering the effect 
 of turbulence in the jets. Since the magnetic field is embedded in the jets so turbulence in jets can cause turbulent magnetic field which
 will affect mostly the synchrotron emission and that can lead to the optical variability while turbulent magnetic field cannot affect the 
 $\gamma$-ray radiation. In other words, $\gamma$-ray emission region could be better aligned along the 
 line of sight, which can lead to a higher Doppler factor of high energy flux, as compared to optical emitting region. 
 \\
 A correlation study between $\gamma$-ray and IR/optical/UV has also been done before for some of the blazar e.g. \citet{Bonning et al. (2009)};
 \citet{Vercellone et al. (2009)}; \citet{Raiteri et al. (2011)}; \citet{Jorstad et al. (2013)}; \citet{Larionov et al. (2013)} and 
 \citet{Cohen et al. (2014)}, where they suggested the co-spatial origin of $\gamma$-ray and IR/optical/UV emission. It is also possible that the
 nature of the correlation between two emitted fluxes changes with epochs and it can be seen as an involvement of different processes and/or 
 different particle population during the high activity. 
 \\
 The right plot of lower panel of Figure 7 shows the correlation between $\gamma$-ray and radio (OVRO; 15 GHz). A lag of 27 days in the radio 
 emission at 15 GHz is noted with DCF peak of 0.84. Since the $\gamma$-ray and optical emission is well correlated with a small time lag which 
 suggests that radio emission also lags with optical by the same amount as with $\gamma$-ray.\\
 Time delay uncovered by DCF analysis can relate to the relative location of the emission region at different wavebands, which depends on the
 physics of the jets and high energy radiation mechanisms. The lag of 27 days in the radio emission with $\gamma$-ray/optical clearly says that
 these two emissions are from two different locations in the jets. 
 The observed time lag between $\gamma$-ray and radio can be used to determine the distance between two emitting regions by using the equation 
 given in \citet{Fuhrmann et al. (2014)}
 \begin{equation}
  \Delta r_{\gamma,r} = \frac{\beta_{app} c \Delta t}{sin \theta},
 \end{equation}
 Where $\theta$ = viewing angle of the source, $\beta_{app}$ = apparent jet speed, and $\Delta$t = observed time lag.
 Using $\Delta$t = 27 days, and $\theta$ = 4.3 degree, $\beta_{app}$ = 16.13 from \citet{Liodakis et al. (2017)}, I 
 found $\Delta r_{\gamma,r}$ $\sim$ 5 pc. This means the radio emitting region is located far away from the AGN central engine.
 It is possible that the high energy and radio emission region have different apparent speed as well as different viewing angle which further
 implies that they have different Doppler factor.
 A similar situation is also observed by \citet{Raiteri et al. (2013)} for BL Lacertae,
 where they found a lag of 120-150 days between $\gamma$-ray/optical to radio and the distance between two emitting region
 in a range of 6.5 to 8.2 pc. \citet{Rani et al. (2014)} have also found a time lag of 82 days between $\gamma$-ray and radio emission for S5 0716+714, the
 distance between two emission region is estimated in the range 2.9 - 4.4 pc (\citet{Rani et al. (2015)},
 for $\beta_{app}$ = 6 - 8 c and viewing angle ($\theta$) = 6 - 9 degree). \\
 \\
 Alternatively, Flares which are delayed and appear late at lower frequencies can be seen as a clear indication of opacity effects, in the context of 
 shock-in-jet model (\citealt{Marscher and Gear (1985)}; \citealt{Valtaoja et al. (1992)}), due to synchrotron self-absorption. 
 A shock is formed close to the core where the jet is optically thick to radio frequencies but transparent to high energy, and a component
 at the core of the jet producing optical/$\gamma$-ray flare, propagates along the jets, and after sometime jet becomes optically thin 
 to detect the radio flare.
\section{Conclusions} 
During the end of 2017 blazar Ton 599 went into a long flaring episode throughout the entire electromagnetic spectrum. Flaring was first reported
in $\gamma$-ray followed by the other wavebands. A long delay in the radio flare was observed by OVRO at 15 GHz as seen from Figure 2.
Ton 599 was not much variable in X-ray but its variability can be seen in $\gamma$-ray and UV/optical. In $\gamma$-ray, during the flaring episode a 
maximum flux 12.63$\times$10$^{-7}$ was noticed with photon index 1.81 and a clear brighter and harder spectral behavior is seen (Figure 4).
Large variations in DoP and PA are seen during the flaring period, which can be explained by the shock-in-jet model.
Almost all the peaks of the flare show symmetric profile. The rise and decay time of one of the peak is found to be 2.22$\pm$0.14 and 2.30$\pm$0.13 days. 
Two 42 GeV of photons are detected during the flaring period with a probability of 99.7$\%$ and 99.8$\%$.
For the $\gamma$-rays the size of the emission region is estimated as 1.88$\times$10$^{16}$ cm by
using 1-day as the fastest variability time and the location of the emission region is found to be at the outer edge of the BLR. 
Gamma-ray SED for pre-flare and flare are fitted with four spectral models PL, LP, PLEC, and BPL. For a flare, PLEC gives a better fit to the SED data
points over LP and BPL. A break in the $\gamma$-ray spectrum at 1.11 GeV is seen, which suggest the peak of the IC mechanism lies in the LAT energy
band and the shape of the photon spectrum likely reflects the distribution of emitting electrons.
Ton 599 has shown a trend of high variability with increasing energy. A strong correlation has been seen
between $\gamma$-X-ray, $\gamma$-UV, $\gamma$-Optical, and $\gamma$-radio (15 GHz). A good correlation with the lag of a few days suggests the 
$\gamma$-ray and optical/UV are co-spatial. On the other hand, a lag of 27 days has been observed between $\gamma$-ray and radio (15 GHz) emission, 
suggesting the presence of two different emission zones. The separation between these two emissions region is estimated as $\sim$ 5 pc.
Detailed gamma and radio observations are needed to probe the two different emission region and
a multi-wavelength spectral energy distributions (SEDs) analysis is also required for better constraints on the different emission mechanisms that
are taking place in the jets of blazar Ton 599.\\
\\
\textbf{Acknowledgements :} Author thanks the referee for valuable comments to improve the paper and also thanks to Simran Singh for proof reading.
This work has made use of public \textit{Fermi} data obtained from FSSC.
This research has also made use of XRT data analysis software (XRTDAS) developed by ASI science data center, Italy. Archival data from the Steward
observatory is used in this research. This research has made use of data from the OVRO 40-m monitoring program \citep{Richards et al. (2011)} which
is supported in part by NASA grants NNX08AW31G, NNX11A043G, and NNX14AQ89G and NSF grants AST-0808050 and AST-1109911.

\bibliographystyle{plain}

\end{document}